\documentclass[prd,aps,a4paper,nofootinbib,twocolumn]{revtex4}  

\newif\ifusesec
\usesectrue  
   
\usepackage{graphicx} 
\usepackage{mathrsfs}
\usepackage{amsmath,amsfonts,amssymb}
\usepackage{multirow}

\newcommand{\beq}{\begin{equation}}
\newcommand{\eeq}{\end{equation}}
\newcommand{\bea}{\begin{eqnarray}}
\newcommand{\eea}{\end{eqnarray}}

\begin{document}

\title{Gravitational bremsstrahlung waveform at the eighth post-Minkowskian order in the extreme-mass-ratio limit}

\author{Andrea Geralico}
\affiliation{
Istituto per le Applicazioni del Calcolo ``M. Picone,'' CNR, I-00185 Rome, Italy
}

\date{\today}

\begin{abstract}
The gravitational waveform generated by the scattering of two nonspinning bodies is computed in the frequency domain in the extreme-mass-ratio limit at the eighth post-Minkowskian (PM) order (i.e., $O(G^8)$, or six-loop) and at the fractional sixth post-Newtonian (PN) order.
Previous results at $O(G^4)$ are completed here by computing the 5PM radiated angular momentum as well as the 6PM radiation-reacted scattering angle.
Up to that order the waveform is expressed in terms of few master integrals, with integrands bilinear in (modified) Bessel functions, leading to iterated Bessel functions which can be in turn expressed in terms of Meijer G functions.
Starting from $O(G^5)$ (four-loop) the structure of Fourier integrals becomes quite involved.
In fact, there are several new families of master integrals, which can be shown to satisfy inhomogeneous Bessel equations with master integrals of lower order as sources.
Although limited to the first order in the mass ratio, the results presented here significantly improve the accuracy of the scattering waveform, currently known at the one-loop level from quantum-amplitude-based computations or at the two-loop level (but with 2PN accuracy only) by using the multipolar-post-Minkowskian formalism.
\end{abstract}


\maketitle

\section{Introduction}

The study of the gravitational radiation emitted during the large-impact-parameter scattering of two massive bodies dates back to the pioneering works of Peters \cite{Peters:1970mx}, Kovacs and Thorne \cite{Kovacs:1977uw,Kovacs:1978eu}, and Turner and Will \cite{Turner:1977,Turner:1978zz}.
The bremsstrahlung waveform was obtained there at the leading post-Minkowskian (PM) order by using classical perturbation theory and post-Newtonian (PN) methods.
The results of Kovacs and Thorne are valid for arbitrary values of the velocity (i.e., PN-exact), and were recently revisited by using quantum-field-theory-based techniques developed in high energy physics and effective field theory (EFT) \cite{Jakobsen:2021smu,Mougiakakos:2021ckm,DiVecchia:2021bdo}.

Many powerful methods have been developed in recent years to study the scattering of two massive bodies in a PM framework.
The observable-based formalism of Kosower, Maybee and O'Connell (KMOC) \cite{Kosower:2018adc} allows for establishing a direct relation between on-shell quantum-mechanical scattering amplitudes and classical observables, such as momentum changes and radiative losses.
Within this framework, the waveform is closely related to the five-point amplitude \cite{Cristofoli:2021vyo}, and can be computed by using advanced tools from generalized unitarity and the double copy \cite{Bern:1994zx,Bern:1995db,Britto:2004nc,Bern:2008qj,Bern:2010ue} as well as modern techniques developed for multi-loop computations of Feynman integrals, including integration-by-parts (IBP) identities and differential equations \cite{Chetyrkin:1981qh,Kotikov:1990kg,Remiddi:1997ny,Smirnov:2019qkx}.
Using these tools, the computation of the bremsstrahlung waveform has been extended to the one-loop level \cite{Brandhuber:2023hhy,Herderschee:2023fxh,Elkhidir:2023dco,Georgoudis:2023lgf}. 
This is the state-of-the-art of amplitude-based computations in the observable-based KMOC formalism.

The bremsstrahlung waveform in the frequency domain has been also computed by using the traditional Multipolar-Post-Minkowskian (MPM) formalism \cite{Blanchet:1985sp,Blanchet:1986dk,Blanchet:1987wq,Blanchet:1989ki,Damour:1990gj,Damour:1990ji,Blanchet:1992br}, which combines the PM approximation of the vacuum gravitational field in the weak-field region outside the source with the PN multipole expansion of the source by means of an asymptotic matching procedure. 
The one-loop waveform accurate to the 2.5PN order was obtained in Ref. \cite{Bini:2023fiz}, and initially did not agree with amplitude-based results. The EFT-MPM mismatch was then resolved in Refs. \cite{Georgoudis:2023eke,Georgoudis:2024pdz,Bini:2024rsy} by a suitable reorganization of the amplitude-based expressions.
These results have then been extended to the two-loop order at the 2PN accuracy in Ref. \cite{Bini:2024ijq}, and to the 3.5PN order in Ref. \cite{Bini:2026dvn} (but limited to the quadrupolar part of the waveform). 

In a previous work \cite{Geralico:2026kbm} I computed the frequency-domain bremsstrahlung waveform at the 5PM (three-loop) level with 6PN accuracy to first order in the binary's mass ratio, i.e., within the first-order self-force (1SF) approximation, by using the standard Teukolsky formalism (see, e.g., Ref. \cite{Sasaki:2003xr} for a review).
These results are extended here to the 8PM (six-loop) level, i.e., five PM orders beyond the current knowledge of the EFT waveform, thus providing a test-bed for multiloop calculations in the years to come.
Furthermore, I complete previous analysis of radiative losses by computing the 5PM radiated angular momentum as well as the 6PM radiation-reacted scattering angle.

Units are chosen so that $G=1=c$, unless otherwise specified.
The placeholder $\eta\equiv\frac1c$ is used to keep track of the order of the PN expansion of the various quantities.
The PM expansion is instead an expansion in powers of $G$, or equivalently in powers of the inverse dimensionless angular momentum $\frac1j= \frac{G m_1 m_2}{cJ}$. The latter is related to the impact parameter $b$ by $\frac1j= \frac{G M h}{ b p_\infty}$, where $h=E/Mc^2\equiv\sqrt{1+2\nu(\gamma-1)}$ denotes the dimensionless incoming center-of-mass energy, with $\nu=m_1m_2/M^2$ the symmetric mass ratio, $M=m_1+m_2$ the total mass and $p_\infty=\sqrt{\gamma^2-1}$.
If $m_2>m_1$ the extreme-mass-ratio limit corresponds to $q=\frac{m_1}{m_2}\ll1$, so that one can consistently expand all quantities in powers of $\nu=q+O(q^2)$ at fixed $M$, keeping only terms which are linear in $\nu$.

\section{Hyperbolic-like geodesic motion in a Schwarzschild background}

In the extreme mass-ratio limit the waveform can be computed in the framework of first-order perturbation theory, the smaller mass $m_1$ moving along an hyperboliclike geodesic orbit on the equatorial plane of a Schwarzschild spacetime (with mass $m_2$).

Consider then a Schwarzschild spacetime with line element $ds^2=\bar g_{\alpha\beta}dx^\alpha dx^\beta$ written in standard spherical-like coordinates $(t,r,\theta,\phi)$ as
\beq
ds^2 =-fdt^2 + \frac{1}{f}dr^2 + r^2(d\theta^2+\sin^2\theta d\phi^2)\,,
\eeq
where $f=1-\frac{2m_2}{r}$.
A timelike geodesic orbit on the equatorial plane ($\theta=\frac{\pi}{2}$) has parametric equations $x^\mu =x_p^{\mu}(\tau)$ and 4-velocity $\bar u=\bar u^\alpha\partial_\alpha=\frac{dx_p^\alpha}{d\tau}\partial_\alpha$, such that
\bea
\frac{dt_p}{d\tau}&=&\frac{E}{f(r_p)}\,,\nonumber\\
\frac{dr_p}{d\tau}&=&\epsilon_r\left[E^2-f(r_p)\left(1+\frac{L^2}{r_p^2}\right)\right]^{1/2}\,,\nonumber\\
\frac{d\phi_p}{d\tau}&=&\frac{L}{r_p^2}\,,
\eea
where $\epsilon_r=\pm1$ is a sign indicator keeping track of increasing/decreasing radial coordinate, and $E=-\bar u_t$ and $L=\bar u_\phi$ denote the particle's energy and angular momentum per unit mass, respectively.
For hyperbolic-like orbits $E>1$ and $L>L_{\rm crit}(E)$, where $L_{\rm crit}(E)$ is the critical value of the angular momentum for fixed energy corresponding to capture by the hole (see, e.g., Ref. \cite{Barack:2022pde}).
In this case the radial equation above admits three real roots, $r_1<0$, $r_2$ and $r_{\rm min}$, such that $2m_2<r_2<r_{\rm min}$, 
the latter denoting the closest approach distance.
$E$ and $L$ are related to the initial velocity at infinity $V$ and the impact parameter $b$ by
\beq
E=\frac{1}{\sqrt{1-V^2}}\,, \qquad
L=b\sqrt{E^2-1}=bVE\,,
\eeq 
with $0<V<1$ and $b>b_{\rm crit}=L_{\rm crit}(E)/(VE)$.
Weak-field (i.e., PM) solutions to the geodesic equations correspond to large values of the impact parameter, i.e., $m_2/b\ll1$, at fixed values of the velocity $V$. Small values of the velocity then gives the PN expansion.

It is often convenient to introduce the polar representation of the orbit
\beq
r_p=\frac{Mp}{1+e\cos \chi}\,,
\eeq
where $p$ is the (dimensionless) semi-latus rectum and $e$ the eccentricity, which are related to $E$ and $L$ by
\bea
E&=&\sqrt{\frac{(p-2)^2-4e^2}{p(p-3-e^2)}}\,,\nonumber\\
L&=&\frac{Mp}{\sqrt{p-3-e^2}}\,, 
\eea
with $e>1$ and $p>6+2e$.
The relativistic anomaly $\chi$ takes values in the range $(-\chi_\infty,\chi_\infty)$, with $\chi_\infty={\rm arccos}(-1/e)$, the value $\chi=0$ corresponding to the minimum approach distance $r_{\rm min}=Mp/(1+e)$.
In the weak-field limit the orbital parameters $(p,e)$ are expressed in terms of $(b,V)$ as follows (see Ref. \cite{Barack:2022pde} for details)
\bea
\label{epexp}
e&=&V^2\frac{b}{m_2}\left[1-\frac{4V^2-1+8V^4}{2V^4}\left(\frac{m_2}{b}\right)^2\right.\nonumber\\
&&\left.
-\frac{64V^4-8V^2+256V^6+1}{8V^8}\left(\frac{m_2}{b}\right)^4+O\left(\frac{m_2}{b}\right)^6\right]
\,,\nonumber\\
\frac1p&=&\frac{1}{V^2}\left(\frac{m_2}{b}\right)^2\left[1+\frac{4(1+V^2)}{V^2}\left(\frac{m_2}{b}\right)^2\right.\nonumber\\
&&\left.
+\frac{16(2+4V^2+V^4)}{V^4}\left(\frac{m_2}{b}\right)^4+O\left(\frac{m_2}{b}\right)^6\right]
\,,
\eea

However, for the purposes of the present paper the orbit is more conveniently parametrized in the quasi-Keplerian form 
\beq
r_p=\bar a_r (e \cosh v -1)\,,
\eeq
where $\bar a_r=Mp/(e^2-1)=r_{\rm min}/(e-1)$, and the auxiliary parameter $v\in(-\infty,\infty)$ is related to $\chi$ by
\beq
\chi=2\, {\rm arctan}\left(\sqrt{\frac{e+1}{e-1}}\tanh \frac{v}{2} \right)
\,.
\eeq
The expressions for $t_p(v)$ and $\phi_p(v)$ along the orbit can be found by integrating the equations
\bea
\frac{dt_p}{dv}&=&\frac{Mp^2\sqrt{(p-2)^2-4e^2}}{(e^2-1)^{3/2}}(e\cosh(v)-1)\nonumber\\
&&\times
\left(p-6-2e\frac{e-\cosh(v)}{e\cosh(v)-1}\right)^{-1/2}\nonumber\\
&&\times
\left(p-2-2e\frac{e-\cosh(v)}{e\cosh(v)-1}\right)^{-1}
\,,\nonumber\\
\frac{d\phi_p}{dv}&=&\frac{\sqrt{p(e^2-1)}}{e\cosh(v)-1}\left(p-6-2e\frac{e-\cosh(v)}{e\cosh(v)-1}\right)^{-1/2}
\,.\nonumber\\
\eea
We are interested in PN-expanded solutions, which are formally obtained through the replacements $m_2\to m_2\eta^2$ and $1/p\to\eta^2/p$, where the small parameter $\eta=\frac1{c}$ keeps track of the order of the PN expansion (each power of $\eta$ counts as a half PN order).
The structure of the solution is as follows
\bea
  t_p(v)&=&m_2\left[T_{\rm sh}e\sinh v-T_vv\right.\nonumber\\
&&\left.
+\eta^4\left(T_{\chi}\chi(v)+\sum_{k=1}^\infty T_{\rm sh}^{(k)}\left(\frac{m_2}{r_p(v)}\right)^ke\sinh v\,\eta^{2k}\right)\right]
\,,\nonumber\\
 r_p(v)&=&\frac{Mp}{e^2-1} (e\cosh v-1)
\,,\nonumber\\
 \phi_p(v)&=&\Phi_{\chi}\chi(v)+\sum_{k=1}^\infty \Phi_{\rm sh}^{(k)}\left(\frac{m_2}{r_p(v)}\right)^ke\sinh v\,\eta^{2k}
\,,\nonumber\\
\eea
having chosen initial conditions $t(0)=0=\phi(0)$ at the minimum approach distance $r(0)=r_{\rm min}$.
The first few (dimensionless) coefficients have the following PN expansion
\begin{widetext}
\bea
T_{\rm sh}&=&\frac{p^{3/2}}{(e^2-1)^{3/2}}\left[1-\frac{2}{p^2}(e^2-1)\eta^4-\frac{8}{p^3}(e^2-1)\eta^6+O(\eta^8)\right]
\,,\nonumber\\
T_v&=&\frac{p^{3/2}}{(e^2-1)^{3/2}}\left[1-\frac{3}{p}(e^2-1)\eta^2-\frac{6}{p^2}(e^2-1)\eta^4+\frac{6}{p^3}(e^2-1)(e^2-5)\eta^6+O(\eta^8)\right]
\,,\nonumber\\
T_{\rm \chi}&=&\frac{15}{2}p^{1/2}\left[1+\frac{5}{p}\eta^2+\frac{1}{40p^2}(1218+25e^2)\eta^4+\frac{1}{40p^3}(7422+533e^2)\eta^6+O(\eta^8)\right]
\,,\nonumber\\
T_{\rm sh}^{(1)}&=&\frac{35p^{1/2}}{2(e^2-1)^{1/2}}\left[1+\frac{59}{8p}\eta^2+\frac{1}{40p^2}(2187+52e^2)\eta^4+\frac{1}{2240p^3}(878834+53375e^2)\eta^6+O(\eta^8)\right]
\,,
\eea
and
\bea
\Phi_{\chi}&=&1+\frac{3}{p}\eta^2+\frac{3}{4p^2}(e^2+18)\eta^4+\frac{45}{4p^3}(e^2+6)\eta^6+O(\eta^8)
\,,\nonumber\\
\Phi_{\rm sh}^{(1)}&=&\frac{1}{(e^2-1)^{1/2}}\left[1+\frac{33}{4p}\eta^2+\frac{5}{12p^2}(137+4e^2)\eta^4+\frac{35}{64p^3}(61e^2+678)\eta^6+O(\eta^8)\right]
\,,\nonumber\\
\Phi_{\rm sh}^{(2)}&=&\frac{3}{4(e^2-1)^{1/2}}\left[1+\frac{115}{9p}\eta^2+\frac{35}{48p^2}(158+3e^2)\eta^4+\frac{7}{80p^3}(611e^2+10254)\eta^6+O(\eta^8)\right]
\,,\nonumber\\
\Phi_{\rm sh}^{(3)}&=&\frac{5}{6(e^2-1)^{1/2}}\left[1+\frac{273}{16p}\eta^2+\frac{63}{400p^2}(1197+16e^2)\eta^4+\frac{231}{1600p^3}(11842+501e^2)\eta^6+O(\eta^8)\right]
\,.
\eea
\end{widetext}
Substituting the PM expansion \eqref{epexp} of the orbital parameters then gives the combined PM-PN expansion form of the solutions for the geodesic orbit used in the next sections, further taking the low-velocity limit by introducing the small velocity parameter $p_\infty$ according to $V=\eta p_\infty/\sqrt{1+\eta^2p_\infty^2}$.

\section{Gravitational waveform}

The classical waveform $h_{\mu \nu} = g_{\mu \nu}-\eta_{\mu \nu}$ has the PM expansion 
\bea 
\label{hmunu}
h_{\mu \nu}&=&G h^{\rm 1PM}_{\mu \nu}+ G^2 h^{\rm 2PM \, or\, tree}_{\mu \nu} + G^3 h^{\rm 3PM\, or\,  one-loop}_{\mu \nu} \nonumber\\
&+&  G^4 h^{\rm 4PM \, or\,  two-loop}_{\mu \nu} 
+G^5 h^{\rm 5PM \, or\,  three-loop}_{\mu \nu}\nonumber\\
&+&
G^6 h^{\rm 6PM \, or\,  four-loop}_{\mu \nu}
+G^7 h^{\rm 75PM \, or\,  five-loop}_{\mu \nu}\nonumber\\
&+&
G^8 h^{\rm 8PM \, or\,  six-loop}_{\mu \nu}
+O(G^9)\,.
\eea
The linear-in-$G$ contribution is stationary in the time domain, and yields a zero-frequency contribution to
the frequency-domain waveform.
This contribution is not taken into account in the Fourier analyis below, where only strictly positive frequencies $\omega>0$ are considered.
It is useful to introduce the following reduced complex asymptotic waveform
\beq
W=\frac{c^4}{4 G} \lim_{r\to \infty}(r( h_+ -  i h_\times))\,,
\eeq
where $h_+$ and $h_\times$ denote the two transverse-traceless components.
Its PN expansion at the lowest (non-zero frequency) PM order (i.e., $O(G^1)$, or tree-level) thus starts with the classical quadrupole formula at the (fractional) Newtonian order.

According to the Teukolsky formalism, to first order in the mass ratio the waveform can be constructed from the Weyl scalar $\psi_4$, which is asymptotically related to $h_+$ and $h_\times$ by
\beq
\label{hdef}
\psi_4(r\to\infty)\sim-\frac12\ddot h\,,
\eeq
where $h\equiv h_+-ih_\times$, a dot denoting time derivative.
$\psi_4$ satisfies the Teukolsky equation with spin-weight $s=-2$, and can be decomposed as 
\beq
\label{sep}
\psi_4= \frac1{r^4}\int\frac{d\omega}{2\pi}e^{-i\omega t}\sum_{lm}\,\,R_{lm\omega}(r)\,\, {}_{-2}Y_{lm}(\theta,\phi)\,,
\eeq
where ${}_{s}Y_{lm}(\theta,\phi)$ are spin-weighted spherical harmonics (SWSH).
The radial function $R_{lm\omega}(r)$ satisfies the inhomogeneous Teukolsky equation with source term $T_{lm\omega}(r)$. 
The asymptotic solution representing purely outgoing waves is given by 
\beq
R_{lm\omega }(r\to\infty)\sim Z^\infty_{lm\omega} r^3e^{i\omega r_*}\,,
\eeq
where $r_*$ is the tortoise coordinate, and $Z^\infty_{lm\omega}$ is the amplitude
\beq
\label{Zinf}
Z^\infty_{lm\omega}=\frac{C^{\rm trans}_{lm\omega}}{W_{lm\omega}}\int_{2m_2}^\infty dr\frac{R^{\rm in}_{lm\omega}(r)T_{lm\omega}(r)}{\Delta^2}\,.
\eeq
Here $\Delta=r(r-2m_2)$, $W_{lm\omega}$ denotes the (constant) Wronskian, and $C^{\rm trans}_{lm\omega}$ is the (constant) transmission coefficient.
I refer to Ref. \cite{Sasaki:2003xr} for notation and definition of the various quantities. 
 
The asymptotic form of $\psi_4$ then implies
\bea
\label{h}
h
&=&\frac{4G}{r}\sum_{lm}\int\frac{d\omega}{2\pi}{\mathcal W}_{lm}(\omega)e^{-i\omega(t-r_*)}\,\,{}_{-2}Y_{lm}(\theta,\phi)\,,
\eea
where 
\beq
\label{Wlmdef}
{\mathcal W}_{lm}(\omega)\equiv\frac{Z^\infty_{lm\omega}}{2\omega^2}
\eeq
are the waveform modes in the frequency domain.
The latter are integrals of the type
\beq
\label{Wlmdef2}
{\mathcal W}_{lm}(\omega)=\int dt e^{i(\omega t-m\phi_p(t))}{\mathcal F}_{lm\omega}(r_p(t))\,,
\eeq
with the function ${\mathcal F}_{lm\omega}(r_p(t))$ evaluated at the particle position $r=r_p(t)$, and are computed by using the Mano, Suzuki and Takasugi (MST) \cite{Mano:1996mf,Mano:1996vt} solutions satisfying the retarded boundary conditions of ingoing radiation at the horizon and upgoing at infinity.

The function ${\mathcal F}_{lm\omega}(r_p(t))$ can be formally written as a linear combination of the ingoing radial function and its first derivative evaluated at $r=r_p(t)$
\beq
{\mathcal F}_{lm\omega}(r_p(t))=\left[\alpha_{lm\omega}(r) R^{\rm in}_{lm\omega}(r) + \beta_{lm\omega}(r) \frac{dR^{\rm in}_{lm\omega}(r)}{dr}\right]_{r=r_p(t)}\,,
\eeq
with coefficients $\alpha_{lm\omega}$ and $\beta_{lm\omega}$ which also depend on the $\theta$-part of the spin-weighted spherical harmonic with given $lm$ and its derivative evaluated on the equatorial plane (the $\phi$-part having been factored out).
Expressing then the SWSH as linear combination of (the $\theta$-part of) ordinary spherical harmonics $Y_{lm}(\theta)$ (with the same $l$ and $m$) and their $\theta$-derivative $Y_{lm}'(\theta)$ leads to 
\bea
\alpha_{lm\omega}(r_p(t))&=&Y^*_{lm}\left(\frac{\pi}{2}\right) \alpha^{Y}_{lm\omega}(r_p(t))\nonumber\\
&&
+Y_{lm}'{}^*\left(\frac{\pi}{2}\right) \alpha^{Y'}_{lm\omega}(r_p(t))\nonumber\\
&\equiv&\alpha^{\rm even}_{lm\omega}(r_p(t))+\alpha^{\rm odd}_{lm\omega}(r_p(t))\,,
\eea
and analogously for $\beta_{lm\omega}$, so that one can distinguish in the waveform an even-parity part and an odd-parity part 
\beq
{\mathcal W}_{lm}(\omega)={\mathcal W}^{\rm even}_{lm}(\omega)+{\mathcal W}^{\rm odd}_{lm}(\omega)\,,
\eeq
which are in direct correspondence with the mass-type and current-type radiative multipole moments ${\mathcal U}_{lm}\equiv{\mathcal W}^{\rm even}_{lm}$ and ${\mathcal V}_{lm}\equiv{\mathcal W}^{\rm odd}_{lm}$, respectively.

The frequency-domain rescaled waveform ${\mathcal W}=rh/4G$ can then be written as 
\bea
\label{Wfin}
{\mathcal W}(\omega,\theta,\phi)&=&\sum_{lm}{\mathcal W}_{lm}(\omega)\,{}_{-2}Y_{lm}(\theta,\phi)
\,,\nonumber\\
&=&\sum_{l}\left[\eta^{l-2}{\mathcal U}_l(\omega,\theta,\phi)+\eta^{l-1}{\mathcal V}_l(\omega,\theta,\phi)\right]
\,,
\eea
with
\bea
{\mathcal U}_l(\omega,\theta,\phi) &=&\sum_{m=-l}^l {\mathcal U}_{lm}(\omega)\, {}_{-2}Y_{lm}(\theta,\phi)
\,, \nonumber \\
{\mathcal V}_l(\omega,\theta,\phi) &=&\sum_{m=-l}^l {\mathcal V}_{lm}(\omega)\, {}_{-2}Y_{lm}(\theta,\phi)\,.
\eea

In order to compute the integrals \eqref{Wlmdef2} it is convenient to parametrize the geodesics in the quasi-Keplerian form discussed in the previous section, so that 
\beq
\label{Wlmdef3}
{\mathcal W}_{lm}(\omega)=\int dv \frac{dt_p(v)}{dv}e^{i(\omega t_p(v)-m\phi_p(v))}{\mathcal F}_{lm\omega}(r_p(v))\,,
\eeq
taking then both the PM and PN expansions of the integrand.
It is also convenient to use instead of $\omega$ the following dimensionless frequency variable
\beq
u \equiv \frac{\omega b}{p_\infty}\,,
\eeq
with $u>0$, the PN expansion being taken at a fixed value of $u$.

\section{Structure of Fourier integrals}

The SWSH components of the rescaled waveform at the lowest order $G^1$ only involve linear combinations of the modified Bessel functions $K_0(u)$ and $K_1(u)$ at any PN order, with coefficients given by polynomials in $u$ with degree increasing with the PN order.
At $O(G^2,\eta^2)$ a new transcendental function appears, the exponential $e^{-u}/u$ multiplied by a polynomial in $u$.
The Fourier integrals \eqref{Wlmdef3} indeed all have the form
\beq 
\label{Q}
Q_{\alpha}(u) \equiv \int dT \frac{e^{iu T}}{(1+T^2)^{\alpha}} \,,
\eeq
which can be expressed in terms of modified Bessel $K$ functions for a generic index $\alpha$ 
\beq
\label{Q_integ}
Q_{\alpha}(u)
= \frac{ 2^{\frac{3}{2}-\alpha}\sqrt{\pi}  u^{-\frac{1}{2}+\alpha} }{\Gamma(\alpha)}
  K_{-\frac{1}{2}+\alpha}(u)\,. 
\eeq
When $\alpha$ takes integer values, Eq. \eqref{Q_integ} gives Bessel functions of half-integer orders, which reduce to exponential functions.

Starting at order $G^3$ (two-loop) one gets more complicated Fourier integrals leading to higher transcendental functions of $u$.
Working at the 2PN fractional order, it has been shown in Ref. \cite{Bini:2024ijq} that the following further families of integrals appear 
\bea
\label{all_integrals}
Q_{\alpha}^{\rm at}(u)&=&  \int dT \frac{e^{iu T}}{(1+T^2)^{\alpha}}{\rm arctan}\left(T\right)
\,,\nonumber\\
Q_{\alpha}^{\rm as}(u)&=& \int dT \frac{e^{iu T}}{(1+T^2)^{\alpha}}{\rm arcsinh}(T)
\,,\nonumber\\
Q_{\alpha}^{{\rm as}^2}(u)&=& \int dT \frac{e^{iu T}}{(1+T^2)^{\alpha}}{\rm arcsinh}^2(T)\,,
\eea
which can be reduced to the three master integrals $Q^{\rm as}_1(u)$, $Q^{\rm as^2}_{\frac12}(u)$, $Q^{\rm at}_{\frac12}(u)$, upon using IBP identities.
These integrals are iterated Bessel integrals which admit analytic solutions in terms of Meijer G functions.
Starting from $O(\eta^9)$ the further master integral $Q^{\rm at}_1(u)$ enters, as shown in Ref. \cite{Geralico:2026kbm} (working at 6PN accuracy).

When going to $O(G^4)$ (three-loop) there is only the following further family
\beq
Q_{\alpha}^{\rm at\,as}(u)= \int dT \frac{e^{iu T}}{(1+T^2)^{\alpha}}{\rm arctan}(T)\,{\rm arcsinh}(T)\,,
\eeq
with the only new master integral $Q^{\rm at\,as}_{\frac12}(u)$.

Further increasing the PM-PN accuracy, instead, there are several new families of integrals
\begin{widetext}
\bea
Q_{\alpha}^{\rm at^k}(u)&=&  \int dT \frac{e^{iu T}}{(1+T^2)^{\alpha}}{\rm arctan}^k(T)
\,,\nonumber\\
Q_{\alpha}^{\rm as^k}(u)&=& \int dT \frac{e^{iu T}}{(1+T^2)^{\alpha}}{\rm arcsinh}^k(T)
\,, \nonumber\\
Q_{\alpha}^{\rm ln^k}(u)&=& \int dT \frac{e^{iu T}}{(1+T^2)^{\alpha}}\ln^k(1+T^2)
\,, \nonumber\\
Q_{\alpha}^{\rm at^m as^n}(u)&=& \int dT \frac{e^{iu T}}{(1+T^2)^{\alpha}}{\rm arctan}^m(T)\,{\rm arcsinh}^n(T)
\,, \nonumber\\
Q_{\alpha}^{\rm ln^m at^n}(u)&=& \int dT \frac{e^{iu T}}{(1+T^2)^{\alpha}}\ln^m(1+T^2)\,{\rm arctan}^n(T)
\,, \nonumber\\
Q_{\alpha}^{\rm ln^m as^n}(u)&=& \int dT \frac{e^{iu T}}{(1+T^2)^{\alpha}}\ln^m(1+T^2)\,{\rm arcsinh}^n(T)
\,, \nonumber\\
Q_{\alpha}^{\rm ln^m at^n as^k}(u)&=& \int dT \frac{e^{iu T}}{(1+T^2)^{\alpha}}\ln^m(1+T^2)\,{\rm arctan}^n(T)\,{\rm arcsinh}^k(T)
\,.
\eea
\end{widetext}
For instance, up to $O(G^7,\eta^{12})$ (six-loop) one has $Q_{\alpha}^{\rm at^2}(u)$, $Q_{\alpha}^{\rm at^3}(u)$, $Q_{\alpha}^{\rm as^k}(u)$ with $k=3,\ldots,7$, $Q_{\alpha}^{\rm ln}(u)$, $Q_{\alpha}^{\rm ln^2}(u)$, etc.
For each family one can derive from IBP identities recurrence relations which allow for order reduction, so that at the end only few master integrals remain.
The latter are listed in Table \ref{table1} for each PM order.
In general such integrals cannot be expressed in terms of known special functions. In fact, most of them can be shown to satisfy inhomogeneous Bessel equations with master integrals of lower order as sources, so that the kernels will contain the product of at least three Meijer G functions.
This feature is discussed in Appendix A.


\begin{table*}  
\caption{\label{table1} List of (new) master integrals entering the (rescaled) waveform at each PM order. 
}
\begin{ruledtabular}
\begin{tabular}{ll}
$O(G^3)$ & $Q_{\frac12}^{\rm at}$, 
$Q_{1}^{\rm as}$, $Q_{1}^{\rm at}$ \\ 
$O(G^4)$ & $Q_{\frac12}^{\rm at\,as}$ \\ 
$O(G^5)$ & $Q_{\frac12}^{\rm at\,as^2}$, $Q_{\frac12}^{\rm ln\,at}$, $Q_{\frac12}^{\rm ln\,as^2}$, $Q_{\frac12}^{\rm ln^2}$, $Q_{\frac12}^{\rm at^2}$,
$Q_{1}^{\rm at\,as}$, $Q_{1}^{\rm ln\,as}$, $Q_{1}^{\rm as^3}$\\ 
$O(G^6)$ & $Q_{\frac12}^{\rm ln\,at\,as}$, $Q_{\frac12}^{\rm at^2\,as}$, $Q_{\frac12}^{\rm ln^2\,as}$, $Q_{\frac12}^{\rm ln\,as^3}$, $Q_{\frac12}^{\rm at\,as^3}$, 
$Q_{1}^{\rm at\,as^2}$, $Q_{1}^{\rm ln\,as^2}$, $Q_{1}^{\rm ln\,at}$, $Q_{1}^{\rm at^2}$\\ 
$O(G^7)$ & $Q_{\frac12}^{\rm at^3}$, $Q_{\frac12}^{\rm at^2\,as^2}$, $Q_{\frac12}^{\rm at\,as^4}$,  $Q_{\frac12}^{\rm ln\,as^4}$, $Q_{\frac12}^{\rm ln\,at\,as^2}$, $Q_{\frac12}^{\rm ln^2\,as^2}$, $Q_{\frac12}^{\rm ln\,at^2}$, 
$Q_{1}^{\rm ln\,at\,as}$, $Q_{1}^{\rm at^2\,as}$, $Q_{1}^{\rm at\,as^3}$, $Q_{1}^{\rm ln\,as^3}$, $Q_{1}^{\rm at^3}$, $Q_{1}^{\rm as^5}$\\ 
\end{tabular}
\end{ruledtabular}
\end{table*}

\section{Results}

I computed the waveform modes \eqref{Wlmdef3} through the 7PM level and 6PN order, i.e., $O(G^7,\eta^{12})$.
To reach such PN accuracy I evaluated the terms in the sum up to $l=14$, so that the rescaled waveform \eqref{Wfin} reads 
\bea
\label{Wom_th_phi}
{\mathcal W}(\omega,  \theta,\phi) &=& {\mathcal U}_2(\omega,\theta,\phi)\\
&+& \eta ({\mathcal V}_2(\omega,\theta,\phi) +{\mathcal U}_3(\omega,\theta,\phi))\nonumber\\ 
&+& \eta^2 ({\mathcal V}_3(\omega,\theta,\phi)+{\mathcal U}_4(\omega,\theta,\phi)) \nonumber\\ 
&+& \cdots \nonumber\\ 
&+&  \eta^{12} ({\mathcal V}_{13}(\omega,\theta,\phi)+{\mathcal U}_{14}(\omega,\theta,\phi))
+ O(\eta^{13})\,.  \nonumber
\eea
${\mathcal U}_2$ only needs to be evaluated at the fractional 6PN level ($\eta^{12}$), whereas higher multipoles require less accuracy
\bea 
{\mathcal U}_2 &\sim& (G+G^2+\cdots+G^7)(\eta^0+ \eta^2 +\cdots+\eta^{12}) \nonumber \\
{\mathcal V}_2 \; {\rm and} \; {\mathcal U}_3 &\sim& (G+G^2+\cdots+G^7)(\eta^0+ \eta^2 +\cdots+\eta^{11}) \nonumber \\
&\cdots& \nonumber \\
{\mathcal V}_{13} \; {\rm and}\; {\mathcal U}_{14} &\sim& (G+G^2+\cdots+G^7)(\eta^0)\,.
\eea
The values of the individual $lm$ coefficients $ {\mathcal U}_{lm},  {\mathcal V}_{lm}$ can be easily recovered by using the orthonormality of the SWSH, e.g.,
\beq
{\mathcal U}_{lm}(\omega)=\int \sin\theta d\theta d\phi \, {\mathcal U}_l(\omega,\theta,\phi)\;  {}_{-2}Y_{lm}^*(\theta,\phi)\,.
\eeq

The explicit expression for the rescaled waveform at each PM order is given in an ancillary file in the form
\beq
{\mathcal W}(u,y,z)={\mathcal W}^{G^1}(u,y,z)+\ldots+{\mathcal W}^{G^7}(u,y,z)\,,
\eeq
in terms of the dimensionless frequency variable $u$ and the angular variables $y=e^{i\theta}$ and $z=e^{i\phi}$. 
It is a function of the master integrals listed in Table \ref{table1} and their first derivatives with respect to $u$ as well as of the Bessel K functions of order $0$ and $1$ and their derivatives (up to the sixth order) with respect to the order.

\subsection{Radiated angular momentum at $O(G^5)$}

In a previous work \cite{Geralico:2026kbm} I computed the radiated energy to infinity at $O(G^6)$ by integrating over frequencies the energy flux expressed in terms of the frequency-domain waveform. 
Similarly, the radiated angular momentum is given by
\beq
J_{\rm rad}^\infty=  -\frac{i}{\pi^2}\int_0^{\infty} d\omega \,\omega\int d\Omega[\partial_\phi{\mathcal W}(\omega,\theta,\phi)]{\mathcal W}^*(\omega,\theta,\phi)\,, 
\eeq
where $d\Omega=\sin\theta d\theta d\phi$ is the standard solid angle element.
However, the above solution for ${\mathcal W}$ is not enough to compute $J_{\rm rad}^\infty$, since Eq. \eqref{Wom_th_phi} represents the time-dependent part only ($\omega\not=0$) of the waveform. 
One must add the $O(G^0)$ contribution from static modes, which can be obtained by using the results for the 1SF time-domain waveform of Ref. \cite{Geralico:2025rof} evaluated at the infinite past, i.e., by taking the limit $t\to-\infty$.

The PM expansion of the radiated angular momentum in a two-body scattering process reads
\beq
\frac{ J^{\rm rad}}{J_{\rm cm}} = \nu  \sum_{n=2}^\infty \frac{{ J}_{n}}{j^n} \,, 
\eeq
where $J_{\rm cm} = b \mu p_\infty/h=GM^2\nu j$ denotes the center-of-mass angular momentum.
The coefficients $J_n$ are functions of $\gamma$ and $\nu$, and are mostly known in a PN-expanded form (i.e., as a series expansion in the small PN parameter $p_\infty$).
Their 1SF expansion is then $J_n=J_n^{\rm 1SF}(p_\infty)+O(\nu)$.

The 5PM-1SF radiated angular momentum turns out to be 
\begin{widetext}
\bea
J_5^{\rm 1SF}&=&\pi\left[
12
+\frac{107}{2} p_\infty^2
+\frac{69 \pi ^2}{5} p_\infty^3
-\frac{101219}{1512} p_\infty^4
+\left(\frac{7488}{25}-\frac{1305 \pi ^2}{112}\right) p_\infty^5\right.\nonumber\\
&&
+\left(-\frac{4922}{175} \log \left(\frac{p_\infty}{2}\right)+\frac{46 \pi ^2}{5}-\frac{561803611}{10584000}\right)p_\infty^6
+\left(\frac{18384}{175}-\frac{5083 \pi ^2}{168}\right) p_\infty^7\nonumber\\
&&
+\left(\frac{4587 \log \left(\frac{p_\infty}{2}\right)}{1960}+\frac{1689 \pi ^2}{280}+\frac{149436753397}{6519744000}\right) p_\infty^8
+\left(\frac{667648}{18375}+\frac{2880341 \pi ^2}{1182720}\right) p_\infty^9\nonumber\\
&&
+\left(-\frac{9249167 \log \left(\frac{p_\infty}{2}\right)}{5821200}-\frac{3893 \pi ^2}{5040}-\frac{387475005310307}{16781821056000}\right) p_\infty^{10}
+\left(-\frac{12895978}{385875}-\frac{216366809 \pi ^2}{15375360}\right) p_\infty^{11}\nonumber\\
&&\left.
+\left(-\frac{7403242117 \log \left(\frac{p_\infty}{2}\right)}{3805401600}+\frac{7763549 \pi ^2}{1774080}+\frac{473136933371418821}{4799600822016000}\right) p_\infty^{12}
+O(p_\infty^{13})\right]
\,.\nonumber\\
\eea
The first few terms (up to $O(p_\infty^6)$, second line) agrees with the 3PN-accurate expression of Refs. \cite{Bini:2021gat,Bini:2022enm}.
The remaining terms (from 3.5PN to 6PN, i.e., from $O(p_\infty^7)$ to $O(p_\infty^{12})$) instead are new with this work.

Therefore, the knowledge of the 6PM-1SF radiated energy (given by Eq. (26) of Ref. \cite{Geralico:2026kbm}) allows for the computation of the 6PM-1SF contribution to the radiation-reacted scattering angle (as defined in Eq. (23) of Ref. \cite{Geralico:2025rof}), whose general PM expansion reads 
\beq
\label{chiraddef}
\chi^{\rm rad} = \sum_{n=3}^\infty \frac{ 2\chi^{\rm rad}_{n}}{j^n}\,,
\eeq
with coefficients $\chi^{\rm rad}_n=\nu\chi^{\rm rad,\,1SF}_n+O(\nu^2)$.

The 6PM-1SF radiation-reacted scattering angle then reads
\bea
\chi^{\rm rad,\,1SF}_6&=&
\pi\left[
\frac{85}{6 p_\infty}
+\frac{15679}{120} p_\infty
+\frac{573 \pi^2}{40} p_\infty^2
+\frac{3402881}{8640} p_\infty^3
+\left(\frac{72768}{175}-\frac{32079 \pi^2}{2240}\right)p_\infty^4\right.\nonumber\\
&&
+\left(-\frac{20437}{700}\log \left(\frac{p_\infty}{2}\right)+\frac{191 \pi ^2}{20}+\frac{3473109613}{5292000}\right)p_\infty^5
+\left(\frac{130744}{525}-\frac{237787 \pi^2}{13440}\right)p_\infty^6\nonumber\\
&&
+\left(\frac{231953 \log \left(\frac{p_\infty}{2}\right)}{39200}+\frac{5351 \pi^2}{1120}+\frac{5550212254333}{26078976000}\right)p_\infty^7
+\left(\frac{9358364}{67375}-\frac{164760751 \pi^2}{4730880}\right) p_\infty^8\nonumber\\
&&
+\left(-\frac{23440253 \log \left(\frac{p_\infty}{2}\right)}{950400}+\frac{4403 \pi^2}{720}+\frac{840587164901563}{19179224064000}\right) p_\infty^9
+\left(\frac{1407238667 \pi^2}{49201152}-\frac{5897279363}{11036025}\right) p_\infty^{10}\nonumber\\
&&\left.
+\left(\frac{6227080097369 \log \left(\frac{p_\infty}{2}\right)}{106551244800}-\frac{82702951 \pi^2}{7096320}-\frac{1012174041611073217}{7679361315225600}\right) p_\infty^{11}
+O(p_\infty^{12})\right]
\,.\nonumber\\
\eea
\end{widetext}

\section{Concluding remarks}

The gravitational bremsstrahlung waveform for nonspinning bodies has been computed in the frequency domain in the framework of first-order self-force theory, i.e., to the first order in the mass ratio, at the 8PM (or six-loop) order with fractional 6PN accuracy, extending previous (5PM) results.
Starting at the 6PM order the structure of Fourier integrals becomes more and more involved, with an increasing number of master integrals upon reduction. In addition, most of them do not admit a closed analytical form in terms of known special functions. 

The results presented here are far beyond the present knowledge of the scattering waveform obtained by either amplitude-based methods (3PM, or one-loop) or the standard MPM formalism (4PM, or two-loop, with 2PN accuracy), even if valid in the extreme-mass-ratio limit only, thus providing a benchmark for ongoing multiloop calculations.

Finally, a previous analysis of radiative losses at the 5PM order has been completed here by computing the 5PM radiated angular momentum as well as the 6PM radiation-reacted scattering angle, which should serve as useful checks for similar calculations by other methods.

\section*{Acknowledgments}

I would like to thank Thibault Damour and Donato Bini for valuable comments. 
I'm grateful to the Istituto per le Applicazioni del Calcolo ``M. Picone,'' CNR, for past support and hospitality during the development of this project.

\appendix

\section{Master integrals}


The only master integrals entering the rescaled waveform at $O(G^3)$ are $Q_{\frac12}^{\rm at}(u)$, $Q_{1}^{\rm as}(u)$, $Q_{1}^{\rm at}(u)$ and $Q_{\frac12}^{\rm as^2}(u)$.
The latter can be expressed in terms of the second derivative of $K_\nu(u)$ with respect to the order $\nu$ evaluated at $\nu=0$ as 
\beq
\label{Qas2_12_sol}
Q_{\frac12}^{\rm as^2}(u)=-\frac{\pi^2}{2}K_0(u)+2\frac{d^2}{d\nu^2} K_\nu(u)\bigg|_{\nu=0}\,,
\eeq
which in turn can be expressed in terms of Meijer G functions \cite{Brychkov:2016} (see also Appendix B of Ref. \cite{Bini:2024ijq}). 
This is a special case of the general formula
\beq
\label{Qask_12_sol}
Q_{\frac12}^{\rm as^k}(u)=2\,(-1)^k\frac{d^k}{d\nu^k} \left(\frac{K_\nu(u)}{i^{\nu}}\right)\bigg|_{\nu=0}\,,
\eeq
which can be easily obtained following the method outlined in Appendix A of Ref. \cite{Bini:2024ijq}.
The remaining integrals are $Q_{\frac12}^{\rm at}(u)=i \widetilde Q_{\frac12}^{\rm at}(u)$, $Q_{1}^{\rm as}(u)=i \widetilde Q_{1}^{\rm as}(u)$, $Q_{1}^{\rm at}(u)=i \widetilde Q_{1}^{\rm at}(u)$, with
\bea
\label{sol_tildeQat12}
\widetilde Q_{\frac12}^{\rm at}(u)&=&\frac{\pi^2}{2}I_0(u) \nonumber\\
&-& 2I_0(u)\frac{\sqrt{\pi }u}{4}  G_{2,4}^{3,1}\left(u^2\bigg|
\begin{array}{c}
 \frac{1}{2},\frac{1}{2} \\
 0,0,0,-\frac{1}{2} \\
\end{array}
\right)\nonumber\\  
&+& 2K_0(u)\frac{u}{4 \sqrt{\pi }}G_{2,4}^{2,2}\left(u^2\bigg|
\begin{array}{c}
 \frac{1}{2},\frac{1}{2} \\
 0,0,-\frac{1}{2},0 \\
\end{array}
\right)
\,,\\
\label{sol_tildeQas1}
\widetilde Q_{1}^{\rm as}(u)&=&\frac{\pi^2}{2}e^{-u}
+\sqrt{\pi}e^{u}G_{2,3}^{3,0}\left(2 u\left|
\begin{array}{c}
 \frac{1}{2},1 \\
 0,0,0 \\
\end{array}
\right.\right)  \nonumber\\
&-&\frac{e^{-u}}{\sqrt{\pi}}G_{2,3}^{3,1}\left(2 u\left|
\begin{array}{c}
 \frac{1}{2},1 \\
 0,0,0 \\
\end{array}
\right.\right)
\,,\\
\label{sol_tildeQat1}
\widetilde Q_{1}^{\rm at}(u)&=&\frac{\pi}{2}\left[e^{u}{\rm Ei}_1(2u)+e^{-u}\left(\gamma+\ln(2)+\ln(u)\right)\right]\,,
\nonumber\\
\eea
whch are real functions of $u$.
In the latter relation ${\rm Ei}_1(x)$ is the exponential integral, which can be in turn expressed in terms of the first derivative of $K_\nu(u)$ with respect to the order evaluated at $\nu=\frac12$ as follows
\beq
\frac{d}{d\nu} K_\nu(x)\bigg|_{\nu=\frac12}=\sqrt{\frac{\pi}{2x}}{\rm Ei}_1(2x)e^x\,.
\eeq

At $O(G^4)$ we have the further integral $Q_{\frac12}^{\rm at\,as}(u)$, which can be written as 
\beq
Q_{\frac12}^{\rm at\,as}(u)= \frac{\pi}{2}\widetilde Q_{\frac12}^{\rm at}(u)+f_{\frac12}^{\rm at\,as}(u)\,,
\eeq
with
\bea
f_{\frac12}^{\rm at\,as}(u)&=&\pi  K_0(u) \left[u \, _3F_3\left(1,1,\frac{3}{2};2,2,2;-2 u\right)\right.\nonumber\\
&&\left.
-\log (u)-\gamma +\log (2)\right]\nonumber\\
&&
-\pi ^{3/2} I_0(u) G_{2,3}^{3,0}\left(2
   u\left|
\begin{array}{c}
 \frac{1}{2},1 \\
 0,0,0 \\
\end{array}
\right.\right)\,.
\eea

Starting at $O(G^5)$ there are several new master integrals, which in general cannot be computed analytically.
For instance, the integrals $Q_{\frac12}^{\rm as^4}(u)$ and $Q_{\frac12}^{\rm as^6}(u)$, Eq. \eqref{Qask_12_sol}, contain the fourth and sixth derivatives of $K_\nu(u)$ with respect to the order, which are not explicitly known in closed analytical form.

Some integrals can be easily computed, e.g., 
\beq
Q_{1}^{\rm at^2}(u)= \widetilde Q_{1}^{\rm at}(u)[\gamma+\ln(2)+\ln(u)]+f_{1}^{\rm at^2}(u)\,,
\eeq
with
\bea
f_{1}^{\rm at^2}(u)&=&-\frac{\pi}{24}\left[\pi^2+6(\gamma+\ln(2)+\ln(u))^2\right]\left(2e^u-e^{-u}\right)\nonumber\\
&&
+2 \pi  e^u u \, _3F_3(1,1,1;2,2,2;-2 u)\nonumber\\
&&
+\frac{1}{2} \pi  e^{-u} G_{2,3}^{3,1}\left(2 u\left|
\begin{array}{c}
 0,1 \\
 0,0,0 \\
\end{array}
\right.\right)\,.
\eea

However, most integrals cannot be explicitly evaluated, since they satisfy inhomogeneous Bessel equations with master integrals of lower order as sources.
Consider for instance $Q_{\frac12}^{\rm at^2}(u)$, which is a solution of the following second order differential equation
\beq
u\frac{d^2}{du^2}Q_{\frac12}^{\rm at^2}(u) +\frac{d}{du}Q_{\frac12}^{\rm at^2}(u)-uQ_{\frac12}^{\rm at^2}(u)=2\widetilde Q_{\frac12}^{\rm at}(u)\,,
\eeq
which is a inhomogeneous modified Bessel differential equation of order $n=0$ with general solution
\bea
\label{sol_Qat212}
Q_{\frac12}^{\rm at^2}(u)&=&c_1K_0(u)+c_2I_0(u)
+2I_0(u)\int^u K_0(x)\widetilde Q_{\frac12}^{\rm at}(x)dx \nonumber\\ 
&-& 2K_0(u)\int^u I_0(x)\widetilde Q_{\frac12}^{\rm at}(x)dx  \,,
\eea
where $\widetilde Q_{\frac12}^{\rm at}(u)$ is given by Eq. \eqref{sol_tildeQat12}.
The integrands in the solution above thus contain the product of two modified Bessel functions and a Meijer G function, or equivalently the product of three Meijer G functions, which in general does not allow for direct evaluation of the corresponding indefinite integral.

\end{document}